\documentclass[12pt]{iopart}
\usepackage{graphicx}
\usepackage{amssymb}
\def\diff{\mathrm d}
\def\kB{k_{\mathrm B}}
\def\Tc{T_{\mathrm c}}
\def\Sm{S_{\mathrm m}}
\def\sm{s_{\mathrm m}}
\def\Q6{Q_{\mathrm 6}}
\newcommand{\bm}[1]{\mbox{\boldmath{$#1$}}}

\begin{document}
\title[The order-disorder transition in colloidal suspensions under shear
flow]{The order-disorder transition in colloidal suspensions under shear flow}
\author{Masamichi J. Miyama and Shin-ichi Sasa}
\address{Department of Pure and Applied Science, University of Tokyo, Komaba, 
Tokyo 153-8902, Japan}
\ead{miyama@jiro.c.u-tokyo.ac.jp and sasa@jiro.c.u-tokyo.ac.jp}

\begin{abstract}
We study the order-disorder transition in colloidal suspensions under shear 
flow by performing Brownian dynamics simulations. We characterize the 
transition in terms of a statistical property of time-dependent maximum value 
of the structure factor. We find that its power spectrum exhibits the power-law 
behaviour only in the ordered phase. The power-law exponent is approximately 
-2 at frequencies greater than the magnitude of the shear rate, 
while the power spectrum 
exhibits the $1 / f$-type fluctuations in the lower frequency regime.
\end{abstract}

\pacs{05.40.-a, 05.70.Fh, 83.80.Hj}

\maketitle

\section{Introduction}
Order-disorder transitions such as solid-liquid transitions and
ferromagnetic transitions are distinctive phenomena in equilibrium
systems. The understanding of the transitions has been a cornerstone for
developments in equilibrium statistical mechanics. Here, the notion
of ``order'' is not restricted to equilibrium systems. For example, one
can modify a system exhibiting the order-disorder transition under equilibrium 
conditions so that it can be observed even under non-equilibrium conditions. 
Since non-equilibrium statistical mechanics has not been established as yet, 
there 
is no systematic understanding of this type of transition from the viewpoint 
of statistical mechanics. Therefore, it is important to
study a typical example related to this question.

As a simple and realistic example, we consider a system consisting of
colloidal particles suspended in a liquid. It is known that this system 
exhibits an order-disorder transition under equilibrium conditions. In 
particular, the so-called ``colloidal crystal'' is observed in the ordered 
phase. This system may be regarded as an ideal model system because one can 
observe such a crystalline structure by using a microscope and also because 
one can control the system as desired.

Recently, a couple of experiments have been reported on colloidal suspensions 
under shear flow \cite{tsuchida:2004, dhont:2005}. Holmqvist et al. 
obtained the phase diagram for colloidal suspensions under stationary 
Couette flow. They observed Bragg reflections of laser light 
and measured the Bragg peak intensity of the first
Debye-Scherrer ring. From the time dependence of the total Bragg peak 
intensity, they determined the crystal growth rate and induction time. The
phase boundary in their phase diagram is determined by extrapolating the 
growth rate and induction time are extrapolated to zero.

Furthermore, there are several reports on numerical experiments of colloidal
suspensions under shear flow. Butler and Harrowell performed Brownian
dynamics simulations of colloidal particles under shear flow 
\cite{harrowell:19951,harrowell:19952}. 
They obtained a phase diagram of temperature versus shear rate. The phase 
boundary of their phase diagram was determined from the long-time average of 
the intensity of the Bragg peak with the wavevector aligned in the 
shear gradient direction. The `order' in their paper was assumed to appear 
when the intensity exceeded half of the intensity value for scattering from a 
body-centred cubic crystal aligned in the direction of the velocity gradient.

These results suggest the existence of the order-disorder transition. Here, let
us recall that crystal is defined as the state with a translational symmetry
breaking. Thus, if we attempt to determine the `crystal' phase, we should
investigate the structure factor $S(k)$ for systems under shear flow. However,
from a simple consideration,  we find that there is no crystal phase in the
rigorous sense, even if the existence of an ordered state is suggested by
several measurements. Now, the objective of this study is to find a useful and
thorough characterization of the ordered state for systems under shear flow. We
attempt this characterization by performing Brownian dynamics simulations of the
colloidal suspensions under shear flow. 

This paper is organized as follows. In section 2, the model that is investigated
is introduced. Further, before discussing non-equilibrium cases in section 3 as
preliminaries, we review the order-disorder (crystal-fluid) transition under
equilibrium conditions and confirm the conventional criteria for the
crystallization. Section 3 comprises the main part of this paper. We
characterize the ordered phase in terms of the time-dependent quantity $\sm(t)$,
which is defined as the first maximum of the structure factor $\hat{s}(k, t)$
for configuration of the particles at time $t$. We find that $\tilde{S}_{\mathrm
m}(\omega)$, which is the power spectrum of $\sm(t)$, exhibits a clear
transition when the temperature is changed. Concretely, in the ordered phase,
the power spectrum exhibits the power-law behaviour, and there are two different
power-law exponent regimes divided by the crossover frequency that is determined
by the shear rate $\dot{\gamma}$: $\tilde{S}_{\mathrm m}(\omega) \simeq
\omega^{-2}$ in the high frequency regime ($\omega \gtrsim 2 \pi \dot{\gamma}$)
and $1/f$-type fluctuations in the low frequency regime ($\omega \lesssim 2 \pi
\dot{\gamma}$). Meanwhile, $\tilde{S}_{\mathrm m}(\omega)$ in the disordered
phase is similar to the white noise spectrum. Future problems are presented in
the final section.

\section{Preliminaries}
We consider a system consisting of $N$ colloidal particles suspended in a fluid
where the stationary planar shear flow is realized. The system is confined to a
cubic cell with a length $L$. The $x$-axis and $z$-axis are chosen to be the
directions of the shear velocity and velocity gradient, respectively. We impose
Lees-Edwards periodic boundary conditions \cite{leeed:1972, smneql:1990} to
avoid peculiarities near the boundaries of the cell. 

We assume that Langevin dynamics can describe the motion of the colloidal
particles. Concretely, the force exerted from the fluid is represented by the
Stokes force and Gaussian noise. In other words, we neglect 
the so-called hydrodynamic effects. Then, the particle positions
$\bm{r}_i(t) = (x_i(t), y_i(t), z_i(t))$, where $1 \leq i \leq N$, 
obey the Langevin equations
\begin{equation}
 \eta \frac{\diff \bm{r}_i}{\diff t} =
  - \sum_{j \neq i} \nabla U(| \bm{r}_i - \bm{r}_j |) 
  + \dot{\gamma} z_i(t) \bm{e}_x 
  + \bm{\xi}_i(t), 
  \label{eq:odlangevin}
\end{equation}
where $\eta$ is a friction coefficient; $\dot{\gamma}$, the shear rate; and 
$\bm{e}_x$, the unit vector that is parallel to the $x$-axis. The variable 
$\bm{\xi}_i(t) = (\xi_i^x(t), \xi_i^y(t), \xi_i^z(t))$ represents the Gaussian 
noise that satisfies
\begin{equation}
 \left< \xi_i^{\alpha}(t) \xi_j^{\beta}(t') \right>
 = 2 \eta \kB T \delta_{ij} \delta^{\alpha \beta} \delta(t - t').
\end{equation}
Here, $\kB$ is the Boltzmann constant and $T$ is the temperature.
The superscripts $\alpha$ and $\beta$ represent the Cartesian components.
Each pair of particles interacts via a screened Yukawa potential 
\begin{eqnarray}
U(r) = \left\{
\begin{array}{ll}
\displaystyle U_0 \sigma
 \left(
  \frac{\exp(-\kappa (r - \sigma))}{r} 
  - \frac{\exp(-\kappa (r_{\mathrm c} - \sigma))}{r_{\mathrm c}}
 \right),
 & \mbox{if $r \leq r_{\mathrm c}$}, \\ 
 0, & \mbox{otherwise}.
\end{array}
\right.
\end{eqnarray}
Here, $r$ is the distance between the particles; $r_{\mathrm c}$, the cutoff 
length that simplifies the calculation in numerical simulations; and $\kappa$, 
a Debye screening parameter.

In this study, all the quantities are converted to dimensionless forms 
by setting $\sigma = U_0 = \eta / \kB T = \kB = 1$. We estimate the 
correspondence between these parameters and those of real experimental systems 
to be $\sigma \sim 10^2$ nm, $U_0 \sim 10 \kB T$ and 
$\kB T / \eta = 10^{-11} {\mathrm m}^{2} {\mathrm s}^{-1}$. 
In our simulation, we assume that $\kappa \sigma = 5.8$, $L/\sigma = 10$, 
$r_{\mathrm c} = 2.5$ and $N\sigma^3/L^3 = 1$. In other words, the Debye 
screening length $\kappa^{-1}$ corresponds to $17$ nm. The typical values of 
the parameters used in our simulations are $T \sim 0.1$ and 
$\dot{\gamma} = 0.001$, and this situation corresponds to experimental systems 
wherein the temperature is 300 K and the shear rate is 1 $\mathrm{s}^{-1}$. 
The systems that have the values are available by laboratory experiments.

In our simulations, we discretized (\ref{eq:odlangevin}) with the time step 
$\Delta t = 0.0025$. Note that in the arguments below, 
$\left< \cdots \right>$ represents the statistical average in the steady 
states. In the calculation we performed, we estimated $\left< A \right>$ to be  
$\int_{t_0}^{t_0 + \tau} dt A(t) / \tau$, where $t_0$ and $\tau$ were chosen 
as that larger than $10^3$, because we had confirmed that the relaxation time 
is approximately $10^2$.

Before considering the behaviour of colloidal suspensions under shear flow, we 
review the transition observed in the system under the equilibrium condition
($\dot{\gamma} = 0$). It is expected that the crystalline arrangement of
colloidal particles is observed in the ordered phase of this system, while they
acquire a random configuration in the disordered phase. In order to 
detect the order-disorder transition, one relies on the definition of a crystal 
according to which the statistical weight for configurations of colloidal
particles breaks 
translational and rotational symmetries. The translational symmetry breaking 
can be quantified by $S(k)$, which is defined as
\begin{equation}
S(k) = \left< \frac{1}{4 \pi} \int_0^{2 \pi} \diff \phi
 \int_0^{\pi} \diff \theta \, \sin\theta \, \frac{1}{N}
 |\tilde{\rho}(\bm{k})|^2 \right>.
\end{equation}
Here, the angles $\theta$ and $\phi$ are defined as 
$\bm{k} = (k \cos \phi \sin \theta, k \sin \phi \sin \theta, k \cos
\theta)$ with $k = |\bm{k}|$, and $\tilde{\rho}(\bm{k})$ is the Fourier 
transform of the number density $\rho(\bm{r})$:
\begin{equation}
 \tilde{\rho}(\bm{k}) = \int_{-\infty}^{\infty} \! \diff^3 \bm{r} \,
  \rho(\bm{r}) \exp(-i \bm{k} \cdot \bm{r})
\end{equation}
with 
\begin{equation}
 \rho(\bm{r}) = \frac{1}{N} \sum_{i=1}^N \delta(\bm{r}-\bm{r}_i).
\end{equation}
When $S(k)$ has a component that is expressed by Dirac's delta function, the
system is assumed to exhibit the translational symmetry breaking. Note that
$S(k)$ can be 
measured experimentally because it is related to the Bragg peak intensity of
the laser light scattering. 

In our simulations, instead of $S(k)$, we use $S_L(k)$ defined by 
\begin{equation}
 S_L(k) = 1 + 4 \pi \rho \int_0^{L/2} \diff r\, 
  (\left< g(r; \{ \bm{r}_i \}) \right> - 1)
  \frac{\sin(kr)}{kr} \frac{\sin(2 \pi r / L)}{2 \pi r / L}, 
 \label{eq:SLk}
\end{equation}
where $\hat{g}(r; \{\bm{r}_i\})$ represents the radial distribution function 
for a given configuration $\{\bm{r}_i\}$, that is, 
\begin{equation}
 \hat{g}(r; \{\bm{r}_i\}) = \frac{1}{4\pi \rho r^2} \frac{\diff n(r)}{\diff r}.
\end{equation}
Here, $\diff n(r)$ is the average number of particles at distances between 
$r$ and $r + \diff r$ from any particle, and its average is taken over all 
the particles. Note that $\left< \hat{g}(r; \{\bm{r}_i\}) \right>$ is equal to 
the standard radial distribution function $g(r)$. It should be noted that 
$g(r)$ is defined in the range where $r \le L / 2$, owing to the periodic 
boundary conditions. The term $\sin(2 \pi r / L) / (2 \pi r / L)$ appended 
in the integrand is the window function to reduce the termination effects 
resulting from the finite upper limit \cite{window:1969,window:2004}. Note 
that $S_L(k)$, as defined above, approaches $S(k)$ in the thermodynamic limit 
$L \rightarrow \infty$.

\begin{figure}[htbp]
 \begin{minipage}{0.5\textwidth}
  \begin{center}
   \includegraphics[width=7cm,clip]{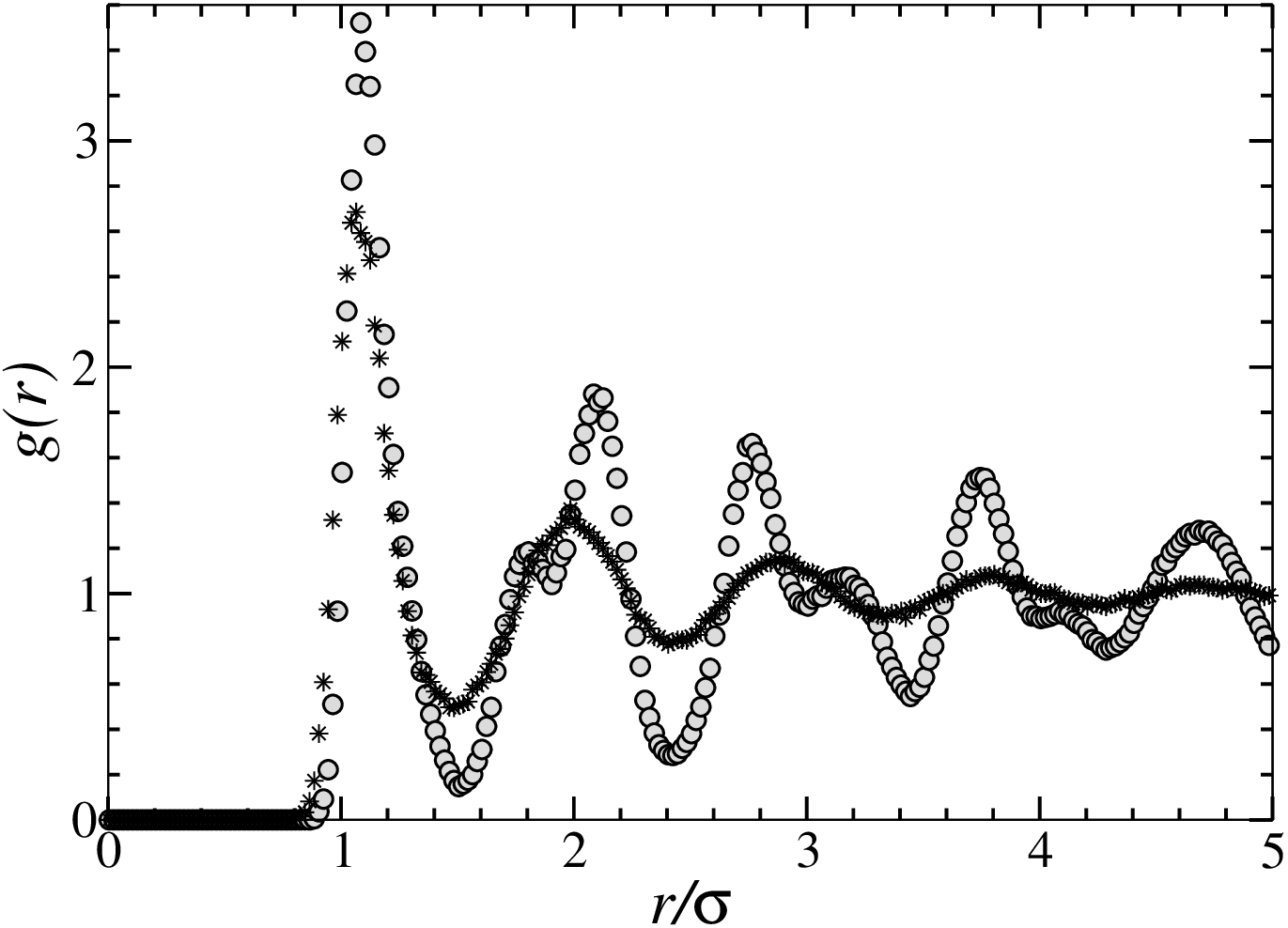}
  \end{center}
  \caption{Plot of $g(r)$ against $r / \sigma$. 
  $T =$ 0.12 ($\opencircle$: open circles) and 0.18 ($\ast$: stars) under equilibrium conditions.}
  \label{fig:greq}
 \end{minipage}
 \begin{minipage}{0.5\textwidth}
  \begin{center}
   \includegraphics[width=7cm,clip]{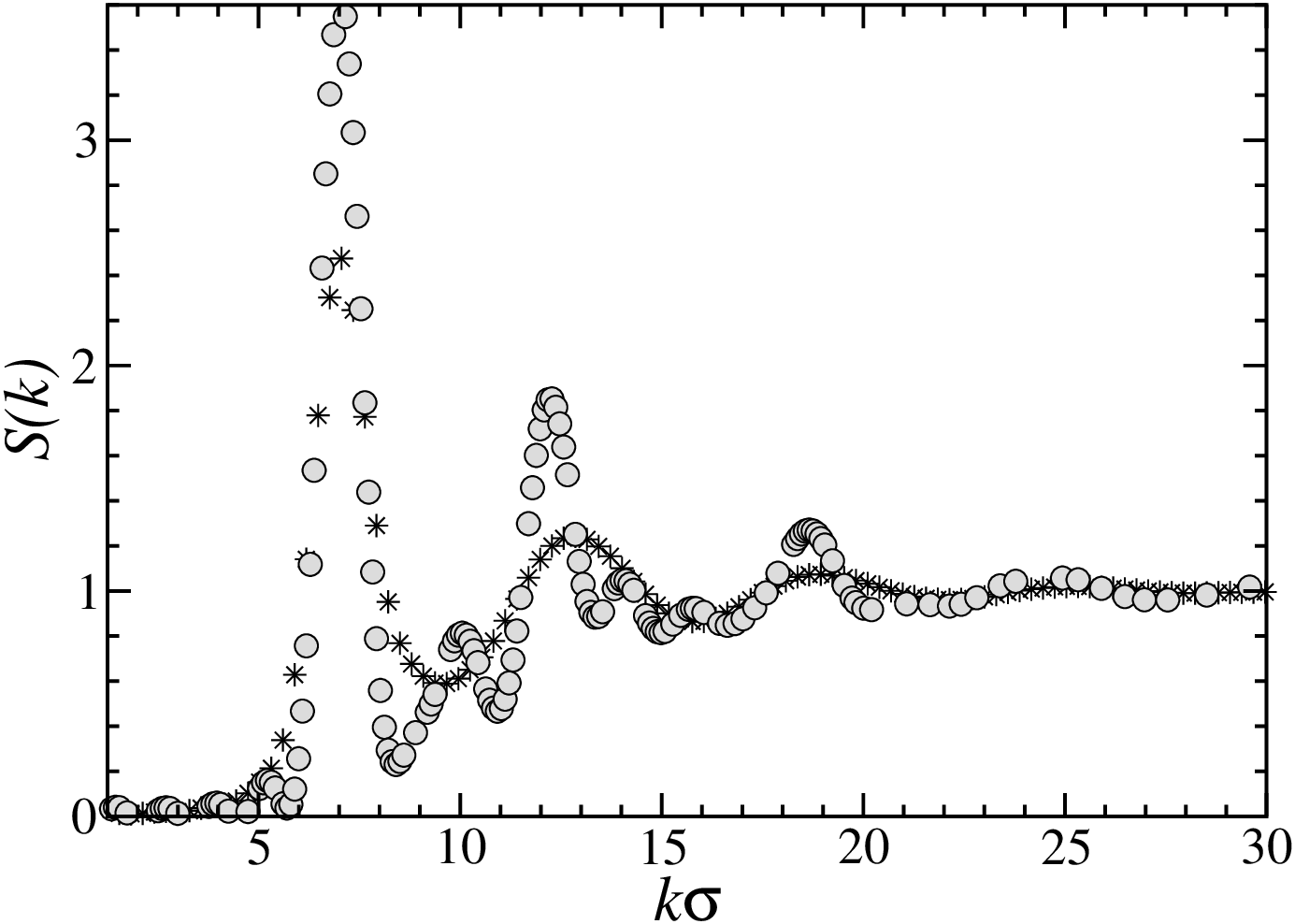}
  \end{center}
  \caption{Plot of $S_L(k)$ against $k\sigma$. 
  $T =$ 0.12 ($\opencircle$: open circles) and 0.18 ($\ast$: stars) under equilibrium conditions..}
  \label{fig:skeq}
 \end{minipage}
\end{figure}

\begin{figure}[htb]
  \begin{minipage}{0.5\hsize}
    \begin{center}
      \includegraphics[width=7cm,clip]{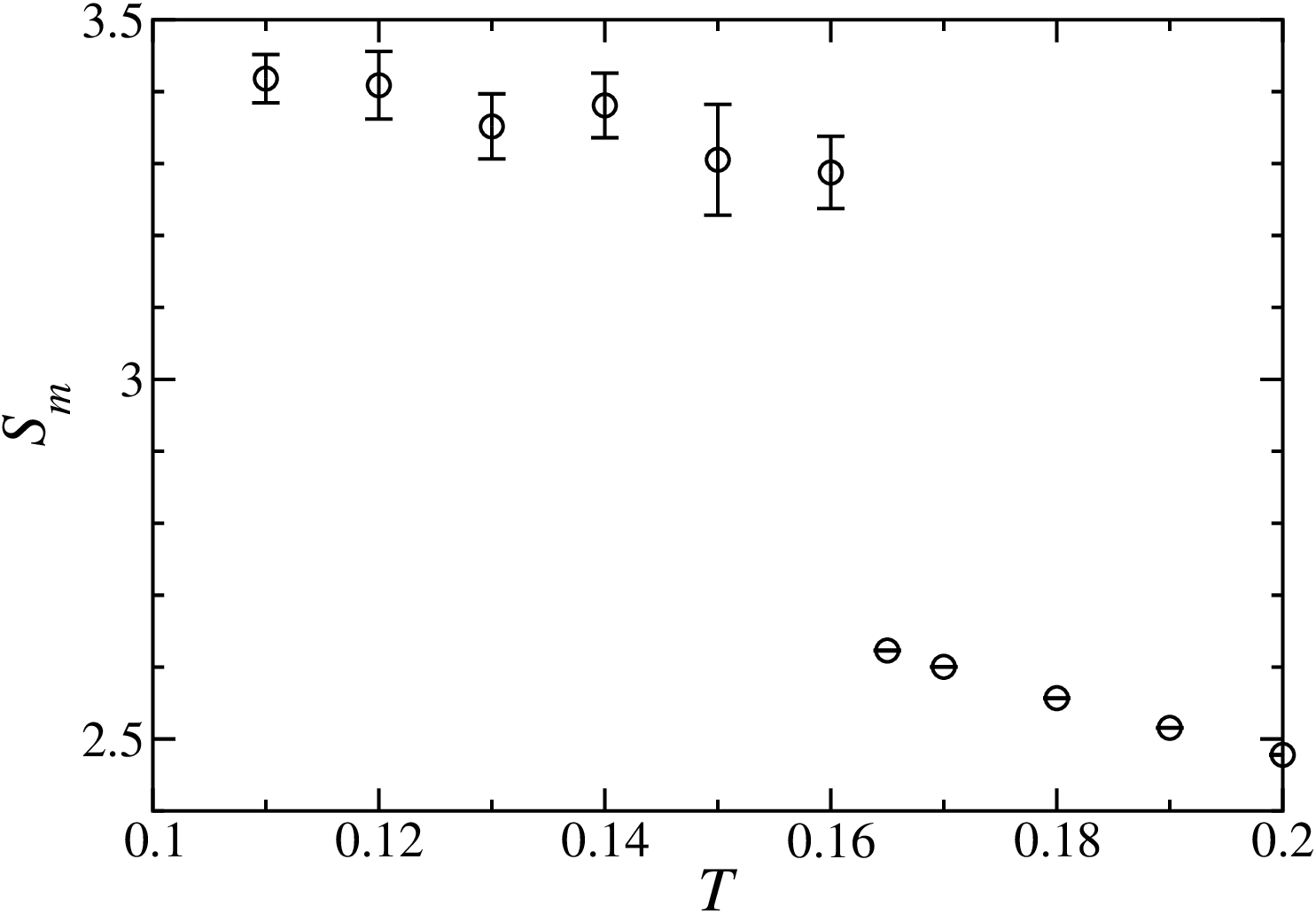}
    \end{center}
    \caption{$\Sm$ as a function of $T$ under the equilibrium condition
   $\dot{\gamma} = 0$. The average values with error bars are displayed using 10
   samples for each temperature.}
    \label{fig:smeq}
  \end{minipage}
  \begin{minipage}{0.5\hsize}
    \begin{center}
      \includegraphics[width=7cm,clip]{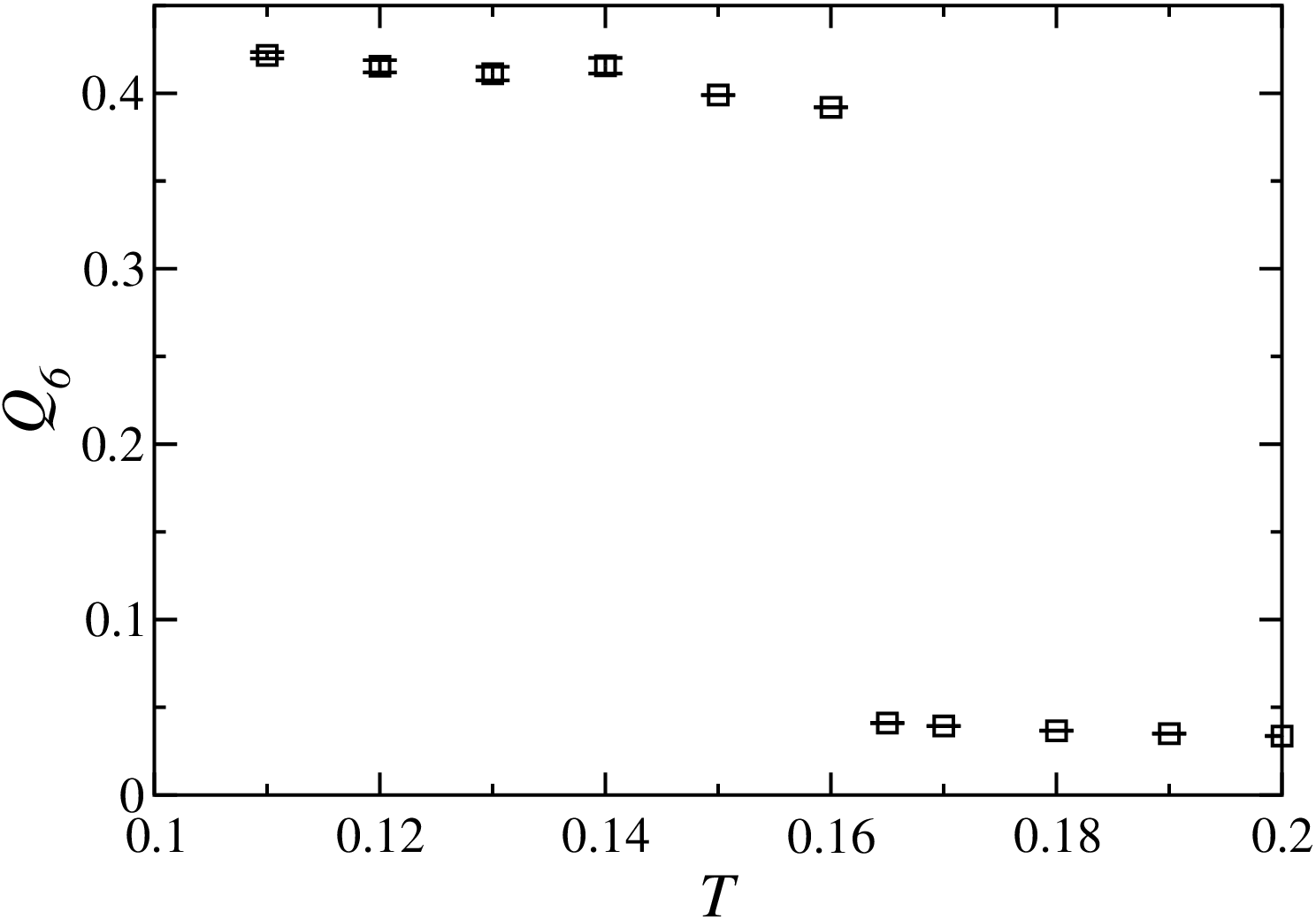}
    \end{center}
    \caption{$\Q6$ as a function of $T$ under the equilibrium condition 
   $\dot{\gamma} = 0$. The average values with error bars are displayed using 10
   samples for each temperature.}
    \label{fig:q6eq}
  \end{minipage}
\end{figure}

As observed in figure \ref{fig:skeq}, we cannot observe infinitely sharp peaks 
owing to finite size effects. However, it is known that there are 
empirical criteria for detecting the transition; one of these criteria is 
Hansen-Verlet's rule \cite{verlet:1969, lowen:2001}. 
According to this rule, a fluid freezes 
when the first maximum value of the static structure factor $\Sm$ exceeds
2.85. This criterion has been tested and validated for various systems
\cite{verlet:1969, lowen:2004}. Indeed, in our system, we find that $\Sm$ 
exhibits a discontinuous jump at the transition temperature $\Tc$ whose value 
is estimated between 0.16 and 0.165, as shown in figure \ref{fig:smeq}. 
Further, 
when the temperature $T$ is lower than the transition temperature $\Tc$, $\Sm$ 
exceeds 2.85, while $\Sm$ is less than 2.85 in the higher temperature 
regime. Thus, we conclude that the order-disorder transition is observed.

As another order parameter for indicating the phase transition, we consider the
bond-orientational order parameter $\Q6$ \cite{srr:1983, frenkel:1992}, which 
is determined by the set of bond vectors  $\{\hat{\bm{r}}_i\}$ as follows:
\begin{equation}
\Q6 = \left< \left(
\frac{4 \pi}{13} \sum^6_{l = -6} |\bar{Q}_{\mathrm 6}^l|^2
\right)^{1/2} \right>,
\end{equation}
with 
\begin{equation}
\bar{Q}_6^l = \frac{1}{N_{\mathrm b}} \sum^{N_{\mathrm b}}_{i = 1}
Y_6^l (\theta(\hat{\bm{r}}_i), \phi(\hat{\bm{r}}_i)).
\end{equation}
Here, each bond vector $\hat{\bm{r}_i}$ corresponds to the relative vector 
between the neighbouring particles; $N_{\mathrm b}$ is the number of bond 
vectors;
$Y_{\mathrm 6}^l$, the spherical harmonics function of degree six; and 
$\theta(\hat{\bm{r}}_i)$ and $\phi(\hat{\bm{r}}_i)$, the polar and 
azimuthal angles of $\hat{\bm{r}}_i$, respectively. Here, we have defined the 
neighbouring particles for a given particle as those within the sphere of  
radius $r_{\mathrm c}$ around the given particle, where $r_{\mathrm c}$ is 
chosen as the first minimum of $g(r)$. The quantity $\Q6$ represents the degree
of breakage of the continuous rotational symmetries, particularly, the 6-fold
rotational symmetry of the configuration of the particles. Its value is 
0.57452 for a face-centred cubic (fcc) crystal, 0.51069 for a body-centred 
cubic (bcc) crystal and 0 for liquids. We show the temperature 
dependence of $\Q6$ in figure \ref{fig:q6eq}. It is observed that $\Q6$ 
decreases
between $T = 0.16$ and $T = 0.165$ in a discontinuous manner. This result is
consistent with that indicated by Hansen-Verlet's rule.

\section{Question and Result}
%
%

\begin{figure}[htb]
 \begin{minipage}{0.5\hsize}
  \begin{center}
   \includegraphics[width=7cm,clip]{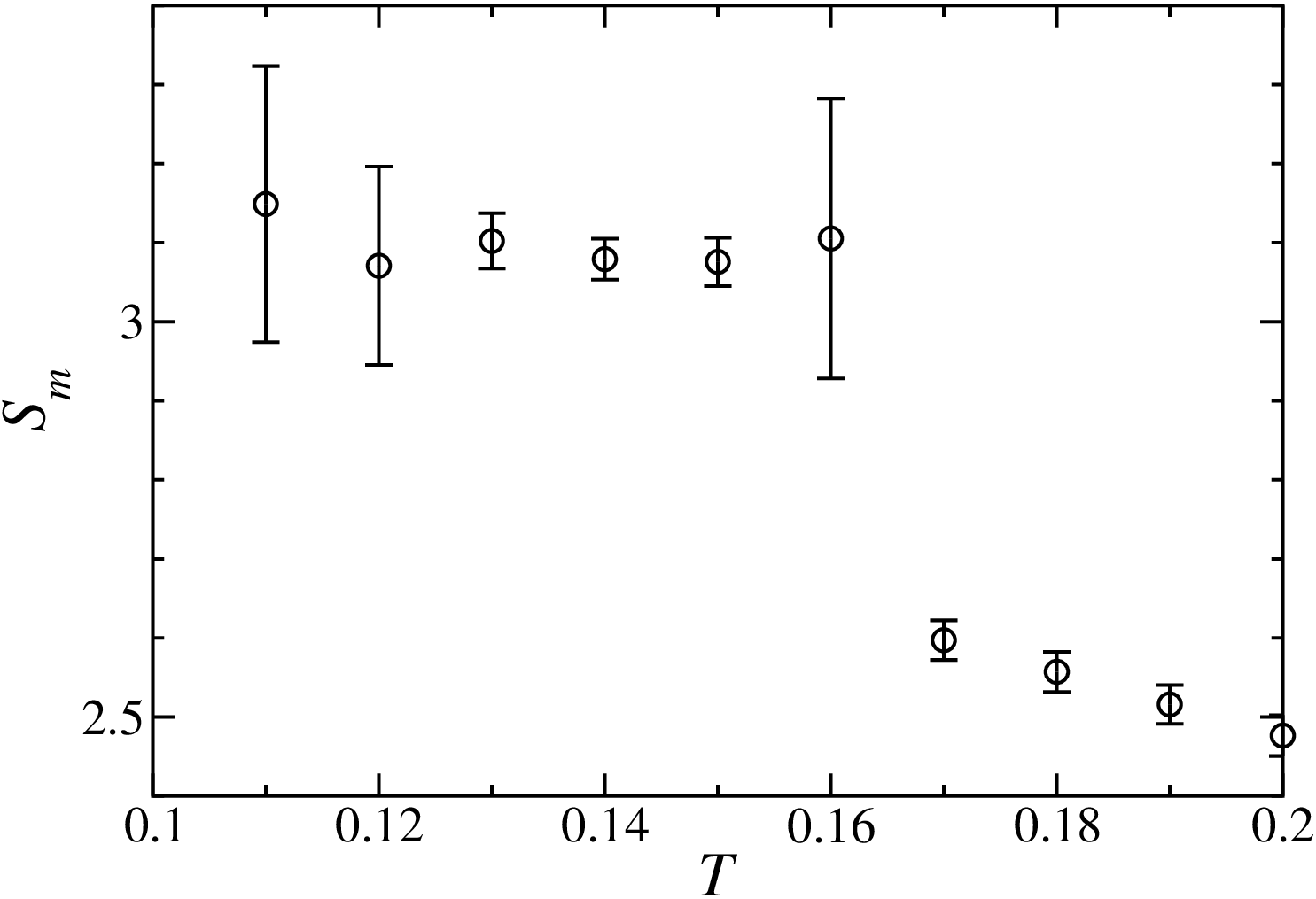}
  \end{center}
  \caption{$\Sm$ as a function of $T$ at $\dot{\gamma} = 0.001$. The average
  values with the error bars are displayed using 10
   samples for each temperature.}
  \label{fig:smneq}
 \end{minipage}
 \begin{minipage}{0.5\hsize}
  \begin{center}
   \includegraphics[width=7cm,clip]{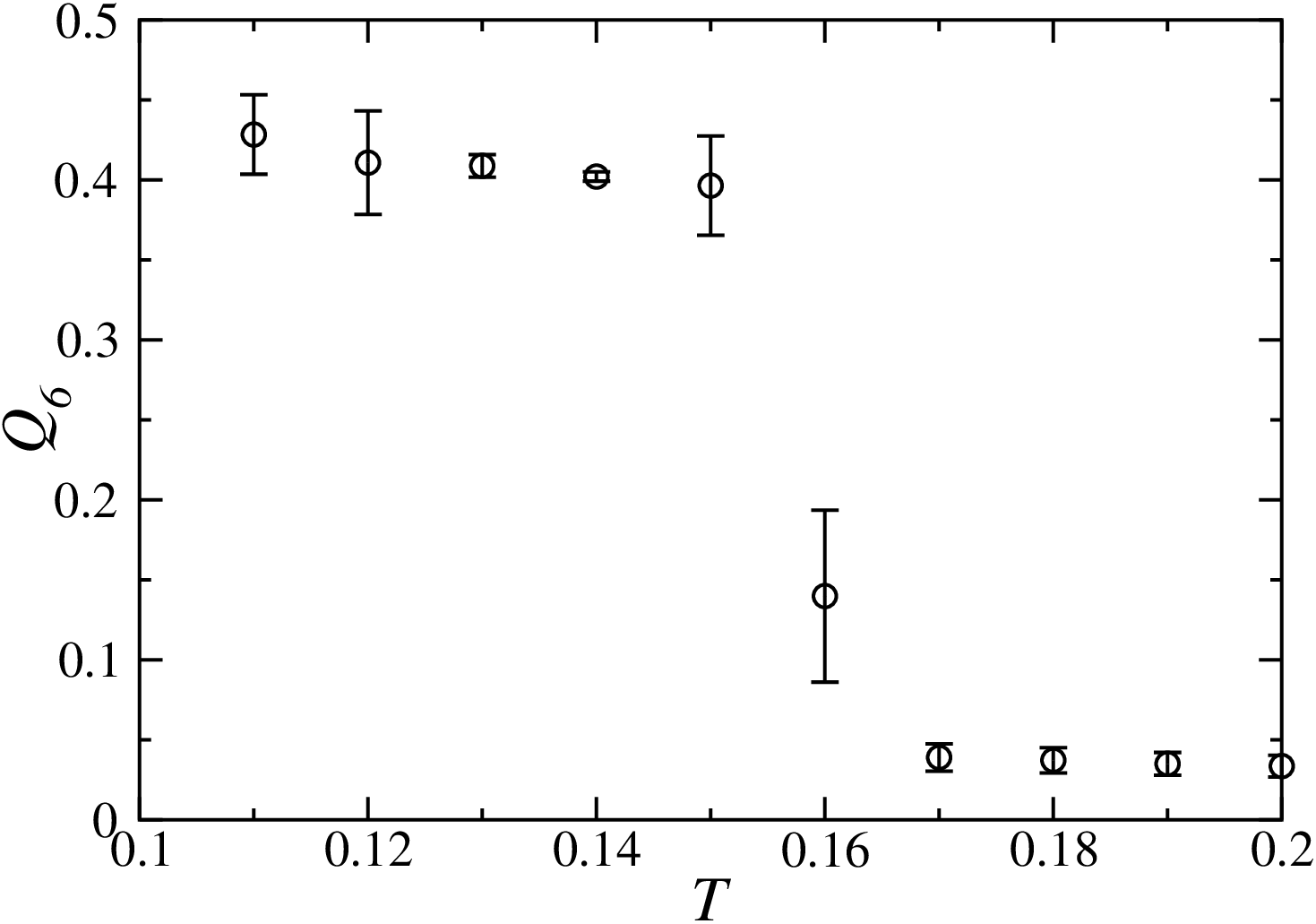}
  \end{center}
  \caption{$\Q6$ as a function of $T$ at $\dot{\gamma} = 0.001$. The average
  values with the error bars are displayed using 10
   samples for each temperature.}
  \label{fig:q6neq}
 \end{minipage}
\end{figure}

Even for colloidal suspensions under shear flow, we can measure $S(k)$ and 
$\Q6$ in a manner identical to that of the equilibrium cases. 
The results with 
$\dot{\gamma} = 0.001$ are displayed in figures \ref{fig:smneq} and 
\ref{fig:q6neq}. 
These graphs are similar to those for the equilibrium system. 
In the temperature regime lower than $T = 0.16$, $\Sm$ exceeds 2.85, 
while it dips from 2.85 in the regime higher than $T = 0.17$. 
Apparently,  this result indicates the existence of an order-disorder 
transition 
in this situation. Similarly, $\Q6$ shows a clear difference between the low
temperature regime and the high temperature regime, as shown in figure
\ref{fig:q6neq}.

%
%

We recall that a crystal is defined 
as the state that breaks the translational symmetry of the statistical
weight for the configurations of the particles. In order to simplify the 
above argument,
we first consider the case $T = 0$ where particles move coherently and 
form two-dimensional crystals in $(x, y)$ planes for a fixed $z$. In 
this state, the translational symmetries in the $x$ and $y$ directions 
are broken for a given $z$. However, the spatial period in the $z$ 
direction is time dependent. Thus, by considering the ensembles generated by
the time evolution, we expect that the translational symmetry in the $z$
direction recovers. Next, we consider the finite temperature cases. A 
translational symmetry is expected to exist in the $z$ direction, and 
according to Mermin-Wagner's theorem, which states that there is no 
translational symmetry 
breaking in two-dimensional systems, we do not expect the symmetry
breaking to occur in the $(x, y)$ planes. Thus, we conclude that there is no 
crystal in colloidal suspensions under shear flow. 

%
%

Although the structure factor does not involve Dirac's delta peak even in the 
thermodynamics limit, figures \ref{fig:smneq} and \ref{fig:q6neq} suggest 
the existence of a discontinuous transition. Therefore, the nature of this 
transition should be different from that in equilibrium systems. In order to
further clarify the above difference, we attempt to study this transition from
another viewpoint.

%
%

\begin{figure}[htb]
\begin{center}
\includegraphics[width=5cm, clip]{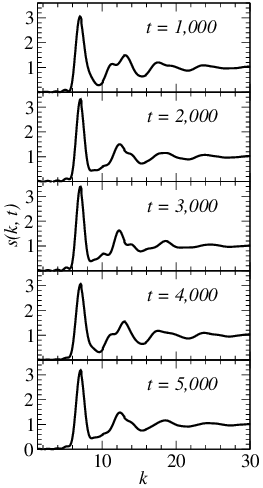}
\includegraphics[width=5cm, clip]{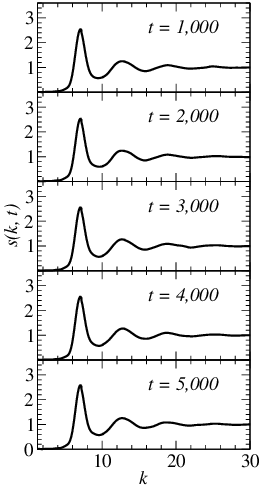}
\end{center}
\caption{Time evolution of the structure factor for $T = 0.14$ (left) and $T =
 0.18$ (right).}
\label{fig:skt}
\end{figure}

In particular, we consider the dynamical features of non-equilibrium systems
possess. First, let us observe the variation of the structure factor with time. 
As an example, we define the time-dependent structure factor $\hat{s}(k, t)$ 
that is determined for each particle configuration $\{\bm{r}_i(t)\}$ at time 
$t$:
\begin{equation}
 \hat{s}(k, t) = 1 + 4 \pi \rho \int_0^{R} \diff r\, 
  (g(r; \{ \bm{r}_i(t) \}) - 1)
  \frac{\sin(kr)}{kr} \frac{\sin(2 \pi r / L)}{2 \pi r / L}.
\end{equation}
Note that $S_L(k)$ defined in (\ref{eq:SLk}) is equal to 
$\left<\hat{s}(k, t)\right>$. In figure \ref{fig:skt}, the results for two 
different cases $T = 0.14$ and $0.18$ are displayed by fixing the shear rate 
to be $\dot{\gamma} = 0.001$. From this figure, it is observed that the maximum 
intensity of $\hat{s}(k, t)$ at $T = 0.14$ varies with time more significantly 
than that at $T=0.18$. Since there appears to be a qualitative difference, 
we focus 
on the time dependence of the first maximum of $\hat{s}(k, t)$ with respect to 
$k$, which is denoted by $\sm(t)$. 

\begin{figure}[htb]
\begin{center}
\includegraphics[width=7cm, clip]{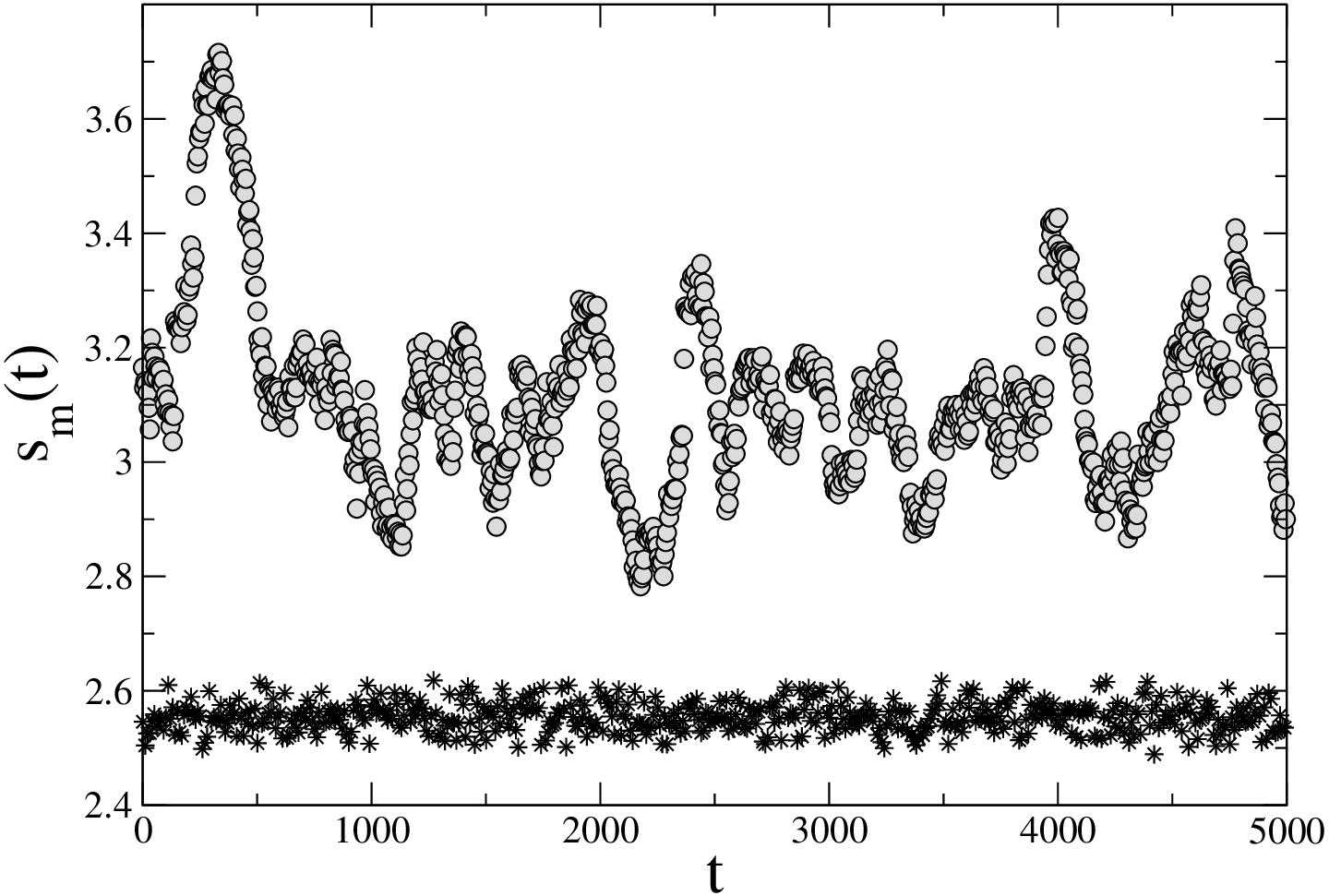}
\end{center}
\caption{$\sm(t)$ for $T = 0.14$ ($\opencircle$: open circles) and $T = 0.18$ ($\ast$: stars).}
\label{fig:smt}
\end{figure}


%
%

\begin{figure}[htb]
\begin{center}
\includegraphics[width=7cm, clip]{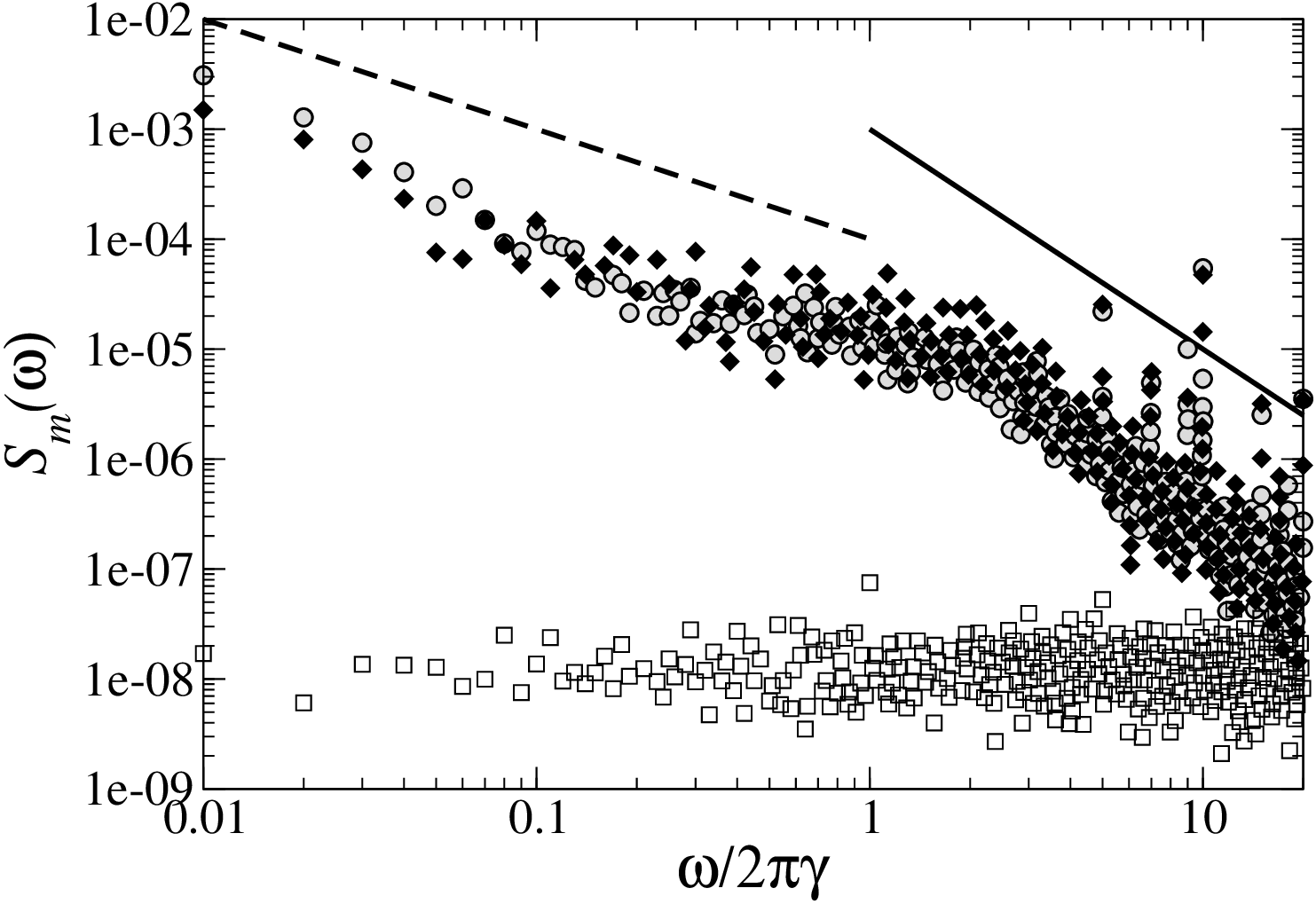}
\end{center}
\caption{Spectra $\tilde{S}_{\mathrm m}(\omega)$ as a function of $\Omega =
 \omega / 2 \pi \dot{\gamma}$. $T = 0.14$ (closed diamonds), 0.16 (open circles) and 
 0.18 (stars). The solid line represents the $\omega^{-2}$ slope, and the 
dashed line denotes the $\omega^{-1}$ slope.} 
\label{fig:skomlog}
\end{figure}

\begin{figure}[htb]
\begin{center}
\includegraphics[width=7cm, clip]{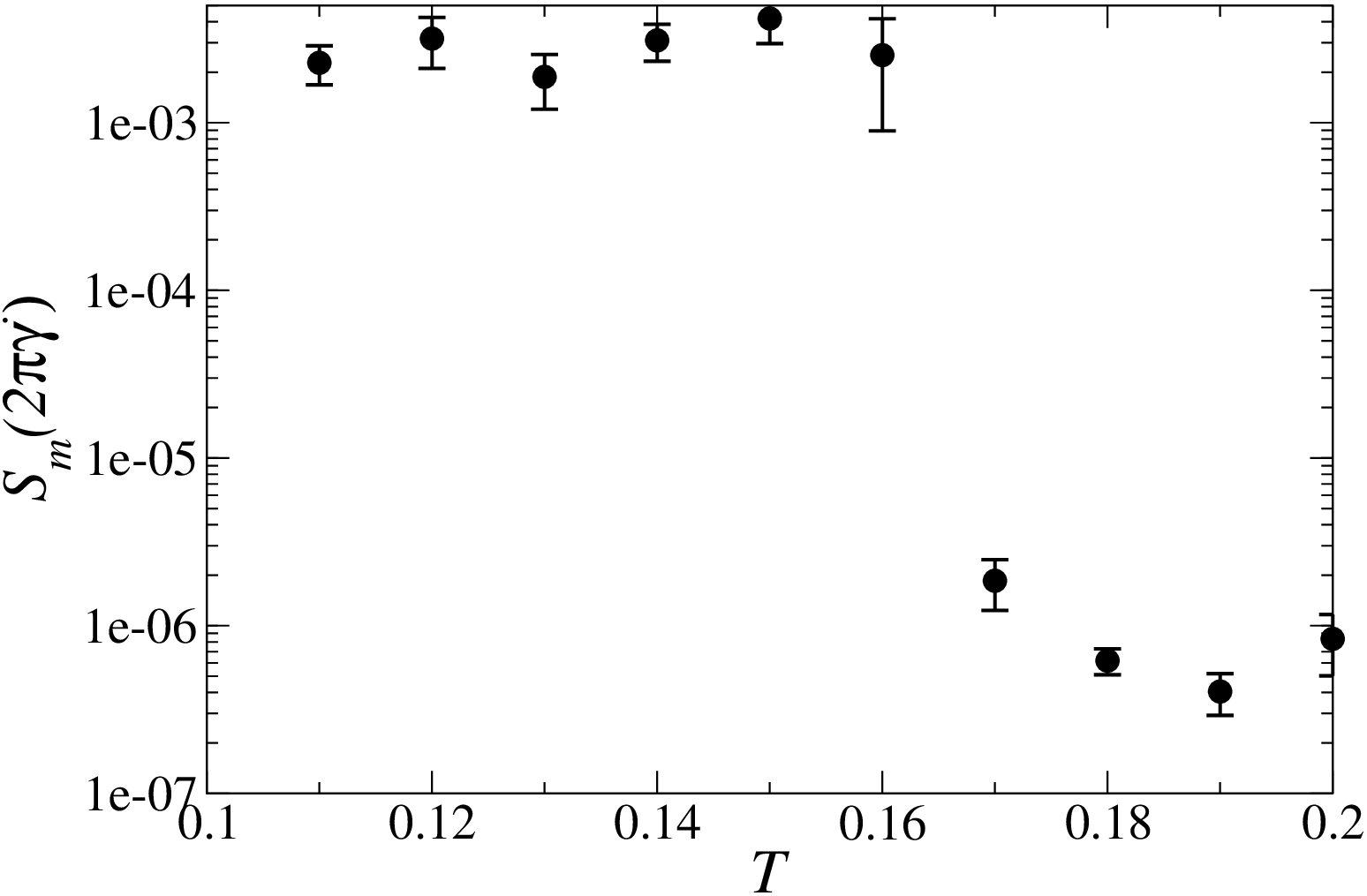}
\end{center}
\caption{$\tilde{S}_m(2\pi \dot{\gamma})$ as a function of $T$.  The average
  values with error bars are determined by using 10
   samples for each temperature.}
\label{fig:smneqgd}
\end{figure}

In figure \ref{fig:smt}, typical data of $\sm(t)$ are displayed. 
It is clearly observed that $\sm(t)$ exhibits a considerably larger fluctuation
in the low temperature case ($T=0.14$)  as compared to that in the high
temperature case ($T=0.18$). 
In order to characterize the difference between the two cases quantitatively, 
we consider the spectra of the fluctuations, which are defined by
\begin{equation}
 \tilde{S}_m(\omega) = \left< \left| \int_{-\infty}^{\infty} \diff t \,
 \sm(t) \exp(-i \omega t) \right|^2 \right>.
\end{equation}
The shapes of the power spectra shown in figure \ref{fig:skomlog} indicate 
a distinct transition at a certain temperature $\Tc$. Indeed, for the 
temperatures $T =0.14$ and 0.16, the power-law behaviour 
$\tilde{S}_m(\omega) \simeq \omega^{-2}$ is observed in the 
frequency regime $\omega \ge 2\pi \dot{\gamma}$. Moreover, focusing on 
the behaviour in the low frequency regime $\omega \le 2\pi \dot{\gamma}$, 
we observe the $1 / f$-type fluctuation. Meanwhile, the spectrum becomes 
flat at the temperature $T=0.18$. Therefore, 
for example, by plotting $\tilde{S}_m(2\pi \dot{\gamma})$ against $T$, 
we can observe a discontinuous transition at $T = \Tc$, as shown in figure 
\ref{fig:smneqgd}.

\section{Concluding Remarks}

The results in this paper motivate us to study the system in more detail.
Before concluding the paper, we address two important future problems. 
%
%
First, the mechanism of the power-law behaviours of $\tilde{S}_m(\omega)$  
should be elucidated on the basis of ``defects'' in the ordered phase. 
After defining the defects in a suitable manner, we may quantitatively 
characterize the destruction of the crystal-like structure and its 
subsequent restoration with time. It is natural to conjecture that the 
power-law behaviour 
$\tilde{S}_m(\omega) \simeq \omega^{-2}$ might be related to the  
generation and annihilation of defects in the ordered phase under shear flow. 
Therefore, by focusing on the elementary processes of defect dynamics, we may 
understand this power-law behaviour. Furthermore, the power-law behaviour 
$\tilde{S}_m(\omega) \simeq \omega^{-1}$ observed in the low frequency regime 
suggests the existence of a more complex mechanism.  
As one possibility, a cooperative phenomenon involving the `defects' may occur. 
In order to explore this possibility, we should investigate the spatial 
correlation of the dynamical events of the defects. 

%
%

The second problem is related to the thermodynamic aspects of the 
order-disorder transition in colloidal dispersions under shear flow.
From the analogy of the order-disorder transition in equilibrium systems, 
we conjecture the existence of a latent heat associated with the order-disorder 
transitions in non-equilibrium systems. However, 
the heat is apparently generated because the system
is in a non-equilibrium steady state. With regard to this 
problem, Oono and Paniconi proposed a remarkable concept in which the excess 
heat plays
a prominent role in a thermodynamic framework for non-equilibrium steady states
\cite{oono:1998}. Indeed, by employing this quantity, the second law of
thermodynamics has been extended 
to transitions between non-equilibrium steady states within the Langevin
description \cite{hatano-sasa:2001}. Thus, we can consider the latent (excess)
heat even for colloidal suspensions under shear flow.  

%
%

In summary, we have characterized the order-disorder transition of colloidal 
suspensions under shear flow by employing dynamical features of the structure 
factor because there exists no `crystal' under the shear flow. Defining a new 
order parameter $\tilde{S}(\omega)$, we have detected the transition 
by measuring the temperature dependence of $\tilde{S}(2 \pi \dot{\gamma})$, 
as shown in figure \ref{fig:smneqgd}. 
We also have found that the discontinuous transition accompanies 
the $1 / f$-type fluctuations in the low frequency regime. By a more detailed 
study of the above-mentioned problems, we hope to gain a thorough understanding 
of the nature of non-equilibrium phase transitions. 

%
%

\ack
We thank K. Kaneko and K. Hukushima for their helpful comments.
This work was supported by a grant (No. 19540394) from the 
Ministry of Education, Science, Sports and Culture of Japan.

\end{document}